\documentclass[aps,10pt,pra,twocolumn,showpacs,amsmath,amssymb, superscriptaddress]{revtex4-1}

\usepackage{graphicx}
\usepackage{color}
\usepackage{transparent}
\usepackage{hyperref}
\usepackage{txfonts}
\usepackage{amsmath}
\usepackage{braket}

\begin{document}

\title{Non local quantum state engineering with the Cooper pair splitter\\beyond the Coulomb blockade regime}

\author{Ehud Amitai}
\affiliation{Department of Physics, University of Basel, Klingelbergstrasse 82, 4056 Basel, Switzerland}
\author{Rakesh P. Tiwari}
\affiliation{Department of Physics, University of Basel, Klingelbergstrasse 82, 4056 Basel, Switzerland}
\author{Stefan Walter}
\affiliation{Institute for Theoretical Physics, University Erlangen-N\"{u}rnberg, Staudtstrasse 7, 91058 Erlangen, Germany}
\author{Thomas L. Schmidt}
\affiliation{Physics and Materials Science Research Unit, University of Luxembourg. L-1511 Luxembourg}
\author{Simon E. Nigg}
\affiliation{Department of Physics, University of Basel, Klingelbergstrasse 82, 4056 Basel, Switzerland}

\begin{abstract}
A Cooper pair splitter consists of two quantum dots side-coupled to a
conventional superconductor. Usually, the quantum dots are assumed to
have a large charging energy compared to the superconducting gap, in
order to suppress processes other than the coherent splitting of Cooper
pairs. In this work, in contrast, we investigate the limit in which the
charging energy is smaller than the superconducting gap. This allows us,
in particular, to study the effect of a Zeeman field comparable to the
charging energy. We find analytically that in this parameter regime the
superconductor mediates an inter-dot tunneling term with a spin symmetry
determined by the Zeeman field. Together with electrostatically tunable
quantum dots, we show that this makes it possible to engineer a spin
triplet state shared between the quantum dots. Compared to previous
works, we thus extend the capabilities of the Cooper pair splitter to
create entangled non local electron pairs. 
\end{abstract}
\pacs{73.23.Hk, 03.67.Bg, 42.70.Qs, 42.50.Dv}
\maketitle

\section{Introduction}

Entanglement~\cite{Schroedinger-1935a} is arguably one of the most fundamental aspects of
quantum mechanics and is an essential resource for emerging quantum technologies. Non local entanglement
manifests itself in correlations between spatially separated parts of
a quantum system that defy any classical explanation. A natural way to explore this phenomenon is by
creating EPR pairs of particles, named after the influential
Einstein-Podolsky-Rosen paper~\cite{Einstein-1935a}, in order to violate Bell's
inequalities~\cite{Bell-1964a, Aspect-1982a, Giustina, Hensen2015, Shalm}. These EPR pairs are the basis for many applications of quantum information theory, such as quantum 
computation~\cite{DiVincenzo13101995}, quantum teleportation~\cite{Bouwmeester-1997a}, and quantum communication~\cite{Ursin}. 

The preparation of EPR pairs of photons is well established in the field
of quantum optics and has already been applied in quantum
teleportation and quantum communication~\cite{Bouwmeester-1997a,
  Ursin}. However, preparing an \emph{electronic} EPR pair has proved
to be rather difficult. Still, a solid state source of electronic EPR
pairs is highly desirable. For example, (on-demand) generation of
electronic EPR pairs would greatly facilitate the construction of quantum repeaters that are
essential ingredients of a future quantum network (quantum
internet)~\cite{Kimble2008}. One promising approach makes use of the natural occurrence of singlet pairs of electrons in
the ground state of conventional s-wave superconductors. By coupling
such a superconductor to two spatially separated quantum dots (QDs),
individual Cooper pairs can split and the two electrons from a pair tunnel to
a different QD each. Because this process is coherent the resulting state
of the two QDs is a non local entangled singlet EPR pair. This process
is dominant if both the superconducting
gap $\Delta$ and the Coulomb
repulsion of electrons on one QD, characterized by the on-site
interaction strength $U>0$, are large compared with the single electron
tunneling rate between the superconductor and the QDs. Such devices
are usually called
{\em Cooper pair splitters} (CPSs) and were first proposed in Refs.~\cite{choi00, recher01, Lesovik-2001a}
and realized experimentally in Refs.~\cite{Hofstetter2009, Herrmann-2010a,
  Das2012}. In these experiments, measurements of the current and
current noise flowing out of the QDs
have confirmed the spatial separation of the electrons from a Cooper pair. Theoretical analysis of the branching currents and their crossed correlations was done in~\cite{Chevallier, Rech}, and the subgap transport was studied in~\cite{Eldridge}. Only recently, measurements of the Josephson current flowing
between superconducting contacts through two parallel QDs have
demonstrated that the pairs are indeed entangled~\cite{Deacon}. 
CPSs can also be used to probe the symmetry of the order parameter in
unconventional superconductors~\cite{tiwari, sothmann15}, as a model system exhibiting unconventional pairing~\cite{Sothmann}, to entangle
mechanical resonators~\cite{Walter}, or to engineer Majorana bound states which are not topologically protected~\cite{Leijnse}.

Typical theoretical treatments of the CPS assume an infinite charging energy for each QD, making it
energetically impossible for two electrons to occupy the same
QD. This is known as the {\em Coulomb blockade approximation} and is
valid as long as the QDs have a relatively large charging energy compared to other
relevant energy scales in the device such as the superconducting gap
and the thermal energy. In the present work we explore the opposite
regime of a small charging energy compared with the superconducting
gap. We show that by leveraging a combination of effects due to Coulomb repulsion, finite
Zeeman magnetic field, and electrostatic tuning of the
system, it is possible to prepare also non local triplet states with zero spin in the
CPS system. This
is particularly interesting for solid state quantum information
processing, where information is encoded in the spin degree of freedom
of electrons trapped in semiconductor QD structures~\cite{DiVincenzo13101995}.

The paper is organized as follows: We begin by summarizing the main results obtained in this paper in Sec.~\ref{Summary}. In Sec.~\ref{System} we describe the model used in this paper for the CPS system. In Sec.~\ref{Effective} we introduce the effective low energy Hamiltonian, obtained for zero temperature and a small charging energy in the QDs compared with the superconducting gap. Sec.~\ref{Triplet_generation} introduces a scheme for the generation of a non local triplet state on the two QDs. In
Appendix~\ref{app:Triplet}, we set up and analytically solve a
simplified model that captures the essential physics of the triplet
generation and in Appendix~\ref{App:SW} we provide details on the
employed Schrieffer-Wolff transformation.


\section{Summary of the main results} \label{Summary}

For simplicity, we restrict ourselves to the
zero temperature limit and consider the coherent dynamics on time
scales that are assumed to be much shorter than the
coherence time of the system. Employing a Schrieffer-Wolff
transformation~\cite{SchriefferWolff}, we integrate out the degrees of freedom of the
superconductor and derive an effective low energy model for
the dynamics of the QDs [see Eqs.~(\ref{Low_Energy_Hamiltonian}) to~(\ref{eq:6})].

As expected, but in contrast to the case of infinite charging energy, this effective low energy model contains a term
that allows two electrons to tunnel to the same QD. This term 
competes
with the Cooper pair splitting process and thus reduces its
efficiency. However, this suppression is of order $\Gamma_0/U$, where
$\Gamma_0$ is the bare Cooper pair splitting rate and can thus be made
small by reducing the tunneling strength between the superconductor
and the QDs at the cost of increasing the duration of singlet
generation $\sim 1/\Gamma_0$.

More interestingly, we also find that for finite on-site Coulomb
repulsion on the QDs, the superconductor
induces an effective inter-dot interaction term. 
In the presence of an (in-plane) magnetic field that lifts the spin
degeneracy via the Zeeman effect by $\Delta_Z$, the spin symmetry of this new term can be
altered and the part which is anti-symmetric under spin exchange can be made to
dominate over the zero-field symmetric part. This effect together with electrostatic tuning of
the QD levels can be used to generate with high fidelity a non local triplet state with zero  spin on the
two QDs. We investigate this triplet generation scheme in detail both
numerically and, within a simplified model also analytically. We find
that in a regime where $\Gamma_0\ll U\ll\Delta_Z\ll\Delta$, the triplet fidelity
that can be achieved is approximately given by
\begin{align}\label{eq:1}
\mathcal{F}_T\approx 1-\left(\frac{U}{\Delta_Z} \right)^2-8\left( \frac{\Gamma_0}{U} \right)^2,
\end{align}
which takes its optimal value $\mathcal{F}_T \approx 1-2^{5/2}\Gamma_0/\Delta_Z$ for $U= 2^{3/4}\sqrt{\Gamma_0\Delta_Z}$.
This simple and intuitive fidelity formula can be used for a quick estimate of parameters
for a given CPS realization. A more general expression for
the fidelity, which relaxes some of the above strong inequality constraints is derived in Appendix~\ref{app:Triplet}.


\section{Description of the physical system and model} \label{System}

\begin{figure}[t]
\includegraphics[width=\columnwidth]{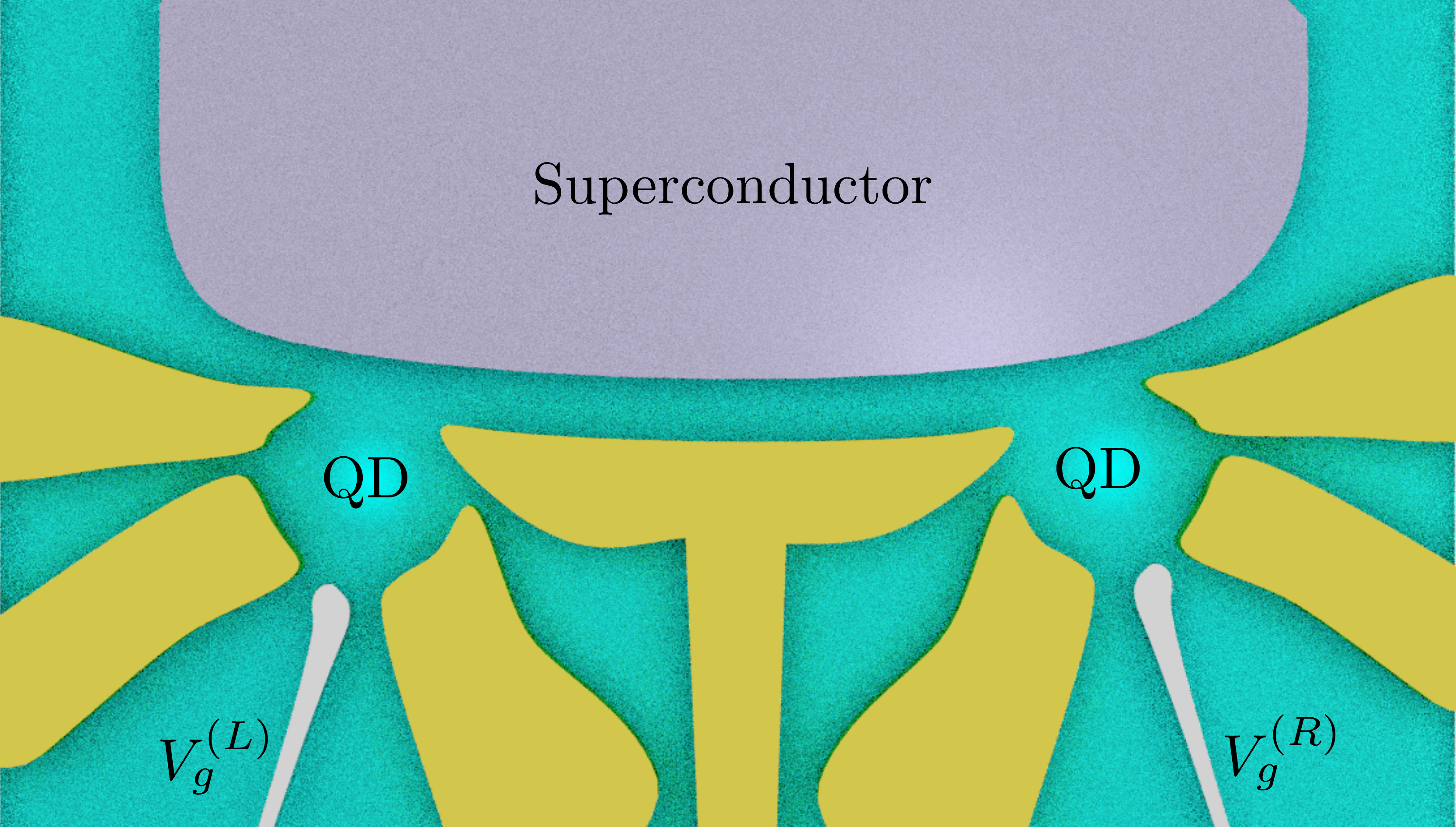}\caption{Schematics of
the Cooper pair splitter system. Such a system can for example be
realized by deposition of a superconductor [wide (purple) structure on
the top] on top of a patterned 2DEG
at the interface of a semiconductor heterostructure. The 2DEG is electrostatically
depleted underneath the superconductor and underneath the gates
defining the quantum dots (yellow structures). Further gates (elongated thin gray structures) can be used to
electrostatically control the potential of the quantum dots.\label{fig:setup}}
\end{figure}

The system we consider is depicted schematically in Fig.~\ref{fig:setup}. It consists
of a conventional BCS superconductor tunnel-coupled to two otherwise
isolated QDs. We assume the coupling is local, which is justified in the limit where the superconducting coherence length is much larger than the distance between the two points on the superconductor from which the electrons tunnel onto the QDs~\cite{recher01, Nigg15, Falci}.  
We assume that only one orbital energy level per QD is
relevant. This approximation essentially requires sufficiently small
QDs with large level spacings. This is typical for QDs from the III-V semiconductors, for which the relevant level would be a state
from the heavy-hole band~\cite{Gywat-2010}. States in the light-hole and conduction
bands can be safely ignored due to their larger detuning from
the chemical potential of the superconductor as compared to the coupling strength.
We consider the zero-temperature limit where
Bogoliubov quasi-particles are absent in the system. The Coulomb
repulsion between two electrons of opposite spins on one QD is
accounted for by the on-site energies $U_L$ (left QD) and $U_R$ (right
QD). The chemical potentials $\mu_L$ and $\mu_R$ of the two QDs can be
tuned electrostatically by a gate. Finally we also allow for an (in-plane)
magnetic field to be applied to the system. This leads to a Zeeman
splitting $\Delta_Z$ of the QD levels. The full
system is modeled by the Hamiltonian
\begin{align}
  H =H_{\rm{BCS}}+H_{\rm QDs}+K \label{Original_Hamiltonian},
\end{align}
where
\begin{align}\label{eq:2}
 H_{\rm{BCS}}=\sum_{\sigma} \sum_k E_k^{} \alpha^\dagger_{k\sigma} \alpha_{k \sigma}^{},
\end{align}
describes the BCS superconductor via the Bogoliubov quasi-particle
operators $\alpha_{k\sigma}$ and energies $E_k$. The Hamiltonians of the
QDs are given by
\begin{align}\label{eq:3}
H_{\rm QDs}=\sum_{\lambda\in\{L,R\}}\sum_{\sigma\in\{+,-\}}\epsilon_{\lambda\sigma}c^\dagger_{\lambda\sigma}c_{\lambda\sigma}+\sum_{\lambda\in\{L,R\}} U_{\lambda}n_{\lambda+}n_{\lambda-}.
\end{align}
Here $c_{\lambda\sigma}$ is a fermionic annihilation operator for an
electron with spin $\sigma$ in QD
$\lambda$. The corresponding number operator is denoted by
$n_{\lambda\sigma}=c_{\lambda\sigma}^{\dagger}c_{\lambda\sigma}$ and the energy levels are given by 
\begin{align}
\epsilon_{\lambda\sigma}=\mu_{\lambda} +\sigma\Delta_Z/2.
\end{align}
Finally, the coupling between the QDs and the superconductor
is given by the tunneling Hamiltonian
\begin{align}
K =\sum_\lambda
  w_{\lambda}\sum_{k\sigma}\left(c_{\lambda\sigma}^{}d^\dagger_{k\sigma}+{\rm
  h.c.}\right),
\end{align}
where the tunnel matrix elements $w_\lambda$ are assumed to be spin
and momentum independent and $d_{k\sigma}$ represents the fermionic
annihilation operator for an electron with energy $\xi_k$ in the
superconductor. In this tunneling Hamiltonian the superconductor is coupled only to the relevant orbital energy level of the QDs (as previously explained).

The electron operators are related to
the Bogoliubov operators in the usual fashion:
\begin{align}
	d_{k+} &= u_k \alpha_{k,+}+v_k \alpha^\dagger_{-k, -},
\\
	d_{-k,-} &= u_k \alpha_{-k, -} - v_k \alpha^\dagger_{k,+},
\end{align}
with $u_k=(1/\sqrt{2})\sqrt{1+\xi_k/E_k}$,
$v_k=(1/\sqrt{2})\sqrt{1-\xi_k/E_k}$ and $E_k=\sqrt{\Delta^2+\xi_k^2}$.

\begin{figure*}[t]
\includegraphics[width=\textwidth]{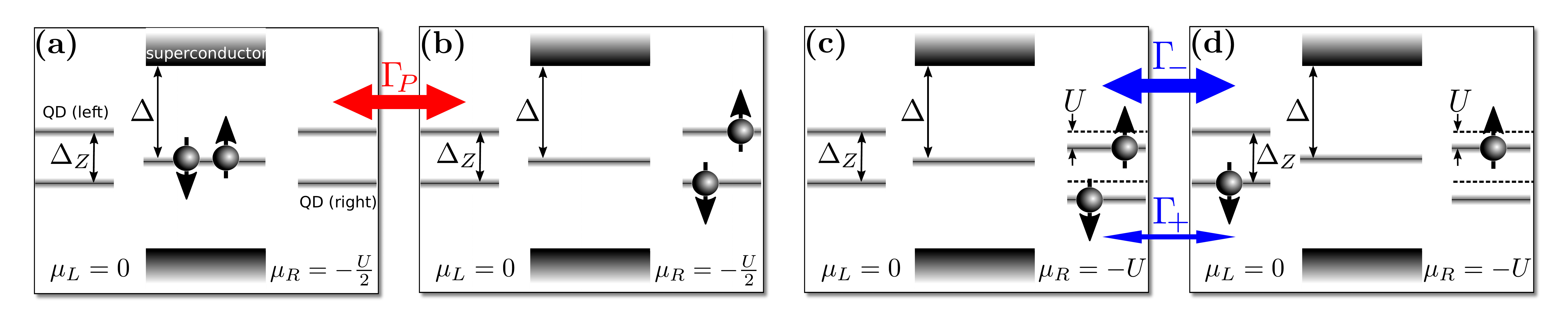}\caption{Schematics
  of the dominant processes for generating the non local triplet
  state
  $\ket{T}=(\ket{+}_L\ket{-}_R+\ket{-}_L\ket{+}_R)/\sqrt{2}$. Panel
  (a) shows the initial state right after switching the potential of
  the right QD to $\mu_R=-U/2$. Both QDs are unoccupied and electrons form
  CPs in the superconductor. A magnetic field lifts the
  spin-degeneracy of the QD levels by $\Delta_Z$. Panel (b) shows the state where
  a CP has been transferred to the right QD at time
  $T_1$. The rate for this process is given by $\Gamma_P$, Eq.~(\ref{Gamma_P}). Panel (c) shows the state right after time
  $T_1$ when the potential of the right QD has been switched to
  $-U$. Taking into account the chemical potential and the charging energy, we see that now the energy levels are shifted by $-U$. Panel (d) shows the state after an electron from the right
  QD (here the down spin electron) has been transferred to the left
  QD at time $T_1+T_2$. This process is driven by the inter-dot
  tunneling term. If $U\ll\Delta_Z/2$, the latter is dominated by the
  spin anti-symmetric term with rate
  $\Gamma_-$, Eq.~(\ref{gamma_-}), as compared with the spin symmetric term with rate $\Gamma_+$, Eq.~(\ref{gamma_+}). The same
  processes but with the spin states interchanged are equally likely and
  their amplitudes add coherently resulting in the generation of a triplet state.~\label{fig:triplet_gen_diagrams}}
\end{figure*}

\section{Effective low energy model} \label{Effective}
Since we are interested in a system where the tunnel coupling between
the QDs and the superconductor is small compared with both the
superconducting gap and the on-site charging energy of the QDs, we
proceed in this section to derive an effective low energy Hamiltonian. This
model will form the basis of our investigation of the CPS beyond the
Coulomb blockade regime.

The first order process described by the tunneling Hamiltonian $K$, is
basically the tunneling of a quasi-particle from the superconductor to
one of the QDs or the conjugate process. However, as we are working in
the limit of zero temperature, quasi-particles are not present. It is
therefore useful to distinguish between the ``high-energy subspace'',
which contains quasi-particle excitations, from the ``low-energy subspace'', which
contains states with no quasi-particles. Transitions between states in
the low-energy subspace can occur via virtual excursions to the
high-energy subspace. This picture suggests the use of the Schrieffer-Wolff (SW)
transformation~\cite{SchriefferWolff}. The SW transformation eliminates the
first order tunneling term from the Hamiltonian, at the expense of
introducing all higher orders. By keeping only the leading order terms
(second order in $w_\lambda$), one obtains an effective low energy
Hamiltonian. This procedure effectively integrates out the degrees of freedom of
the superconductor and allows for a clearer understanding of the CPS dynamics. The details of this
transformation can be found in Appendix~\ref{App:SW}. In order to keep
the following expressions more compact, we consider a
left-right symmetric system, i.e. $w_L=w_R=w$ and $U_L=U_R=U$. The
generalization to asymmetric systems is given in Appendix~\ref{App:SW}. Note that the
chemical potentials of the left and right QDs can still differ from each other. The resulting effective low energy Hamiltonian is given by
\begin{equation}   \label{Low_Energy_Hamiltonian}
	H_{\text{LE}}
=
	H_{\text{QDs}}+H_S+H_P+H_D, 
\end{equation}
where $H_{\rm QDs}$ is given in Eq.~(\ref{eq:3}), and the other terms are given by
\begin{widetext}
\begin{align}
H_S = \frac{\Gamma_0}{2}&
	\left\{
	\left(
			\frac{1}
			{\sqrt{1-\left(\frac{\Delta_Z}{2\Delta}\right)^2}}
			(2-n_{L -}-n_{R+})
	+
			\frac{1}
			{\sqrt{1-\left(\frac{U+\Delta_Z/2}{\Delta}\right)^2}}
			n_{L -}
	+
			\frac{1}
			{\sqrt{1-\left(\frac{U-\Delta_Z/2}{\Delta}\right)^2}}
			n_{R+}
	\right)
	c_{L+}^\dagger c_{R-}^\dagger 
	\right.
\nonumber\\&\left.
	-
	\left(
			\frac{1}
			{\sqrt{1-\left(\frac{\Delta_Z}{2\Delta}\right)^2}}
			(2-n_{L+}-n_{R-})
	+
			\frac{1}
			{\sqrt{1-\left(\frac{U-\Delta_Z/2}{\Delta}\right)^2}}
			n_{L+}
	+
			\frac{1}
			{\sqrt{1-\left(\frac{U+\Delta_Z/2}{\Delta}\right)^2}}
			n_{R-}
	\right)
	c_{L-}^\dagger c_{R+}^\dagger
\right\}+ {\rm h.c.}\label{eq:4}\\
 H_P = \Gamma_P&
		 \left(  c_{R+}^\dagger c_{R -}^\dagger +c_{L+}^\dagger c_{L-}^\dagger \right)
+{\rm h.c.}\label{eq:5}\\
H_D = \frac{\Gamma_0}{2}&\left\{  \left(\frac{1}
	{\sqrt{\left(\frac{\Delta}{U+\Delta_Z/2}\right)^2 - 1}}
	\left(n_{L-}+n_{R-}\right) +
	\frac{1}
	{\sqrt{\left(\frac{2\Delta}{\Delta_Z}\right)^2-1}}
	\left(2-n_{L-}-n_{R-}\right)
	\right)
	c_{L+}^{} c^\dagger_{R+}\right.\nonumber
\\&
	+
	\left.\left(
	\frac{{\rm sgn}(U-\Delta_Z/2)}
	{\sqrt{\left(\frac{\Delta}{U-\Delta_Z/2}\right)^2 - 1}}
	(n_{L+}+n_{R+})-\frac{1}
	{\sqrt{\left(\frac{2\Delta}{\Delta_Z}\right)^2-1}}
	(2-n_{L+}-n_{R+})
	\right)
	c_{L-}^{} c_{R-}^\dagger 
	\right\}+{\rm h.c..}\label{eq:6}
\end{align}
\end{widetext}
Here we have defined the bare resonant Cooper pair splitting rate
$\Gamma_0=\pi\rho_0|w|^2$, where $\rho_0$ is the normal state density
of states at the Fermi energy of the superconductor, and 
\begin{align}\label{Gamma_P}
\Gamma_P=&\frac{\Gamma_0}{2}\left\{\frac{1}
			{\sqrt{1-\left(\frac{\Delta_Z}{2\Delta}\right)^2}}
		+
			\frac{1/2+\frac{1}{\pi}\tan^{-1}\left(\frac{U+\Delta_Z/2}{\sqrt{\Delta^2-\left(U+\Delta_Z/2\right)^2}}\right)}
			{\sqrt{1-\left(\frac{U+\Delta_Z/2}{\Delta}\right)^2}}\right. \nonumber
\\ &\left.
		+
			\frac{1/2+\frac{1}{\pi}\tan^{-1}\left(\frac{U-\Delta_Z/2}{\sqrt{\Delta^2-\left(U-\Delta_Z/2\right)^2}}\right)}
			{\sqrt{1-\left(\frac{U-\Delta_Z/2}{\Delta}\right)^2}}\right\}
\end{align}
is the pair tunneling rate.
Each of the above terms describes a different physical process: $H_S$
describes the Cooper pair splitting, which now depends on the
occupancies of the QDs via the number operators
$n_{\lambda\sigma}$. $H_P$ describes the pair-tunneling to the same
QD. In addition, the superconductor is found to mediate an effective interaction between
the two QDs as described by $H_D$. The latter term has been derived previously in the
infinite $U$ limit~\cite{Nigg15}. However, since double occupancy is strictly
forbidden, this term does not contribute to the dynamics in the latter
case. As we show next, for finite $U$, this term is relevant and can be utilized to
generate a non local triplet state on the two QDs. Equations~(\ref{Low_Energy_Hamiltonian}) to~(\ref{eq:6}) represent
the main technical result of this paper. This effective model is a valid low energy
approximation as long as $\Gamma_0,\Delta_Z/2, U\ll \Delta$. In the limit of $\Delta\rightarrow \infty$, this Hamiltonian agrees with previously known results~\cite{Sothmann}, as shown explicitly in Appendix~\ref{App:SW}.


\section{Triplet generation for finite on-site repulsion and Zeeman field} \label{Triplet_generation}
In this section we present a scheme to generate a non local triplet
state on the two QDs with finite on-site repulsion and in the presence
of a finite Zeeman field.
This scheme is illustrated in Fig.~\ref{fig:triplet_gen_diagrams}. The central ingredient of this scheme is the
inter-dot tunneling term $H_D$ in the regime where
$U<\Delta_Z/2$. In this case, when $H_D$ acts on a state where one
of the QDs is empty while the other is doubly occupied it can be
simplified to
\begin{align}
H_D=\Gamma_+\left(c_{L+}^{}c_{R+}^{\dagger}+c_{L-}^{}c_{R-}^{\dagger}
  \right)+\Gamma_-\left(c_{L+}^{}c_{R+}^{\dagger}-c_{L-}^{}c_{R-}^{\dagger}
  \right)+{\rm h.c.},
\end{align}
with
\begin{align}
\Gamma_+&=\frac{\Gamma_0}{4}\left( \frac{1}{\sqrt{\left( \frac{\Delta}{U+\Delta_Z/2}
  \right)^2-1}}-\frac{1}{\sqrt{\left( \frac{\Delta}{U-\Delta_Z/2}
  \right)^2-1}} \right) \label{gamma_+}
\\
\Gamma_-&=\frac{\Gamma_0}{4}\left( \frac{1}{\sqrt{\left( \frac{\Delta}{U+\Delta_Z/2}
  \right)^2-1}}+\frac{1}{\sqrt{\left( \frac{\Delta}{U-\Delta_Z/2}
  \right)^2-1}}\right)+\left( \frac{\Delta_Z}{2\Delta}
          \right)\frac{\Gamma_S}{2} \label{gamma_-}\\
\Gamma_S&=\frac{\Gamma_0}{\sqrt{1-\left( \frac{\Delta_Z}{2\Delta}
  \right)^2}} \label{gamma_S}.
\end{align}
The terms proportional to $\Gamma_+$ are symmetric under spin exchange and induce non local singlet
pairs while the terms proportional to $\Gamma_-$ are anti-symmetric
under spin exchange and induce non local triplet
pairs. An estimate for the triplet fidelity, given an
initially doubly occupied QD, is derived in
Appendix~\ref{app:Triplet} and is given by
\begin{align} \label{Triplet_fidelity}
\mathcal{F}_T^{\rm ideal}\approx\frac{1}{1+\left(
  \frac{\Gamma_+}{\Gamma_-} \right)^2}\approx 1-\left(\frac{U}{\Delta_Z} \right)^2.
\end{align}
where the last approximation holds if $U\ll\Delta_Z\ll\Delta$.

In order to prepare a state where one QD is empty and the other is
doubly occupied, we take advantage of the Coulomb repulsion and the
gate tunability of the energy levels of the QDs. Specifically, if the
charging energy is such that $U/2\gg\Gamma_0$ and if we
initially detune say the right QD by $\mu_R=-U/2$, then the Cooper pair
splitting term $H_S$ and the inter-dot tunneling term $H_D$ are detuned off resonance and hence suppressed
while the pair tunneling term to the right QD in $H_P=\Gamma_P\left(  c_{R+}^\dagger c_{R -}^\dagger +c_{L+}^\dagger c_{L-}^\dagger \right)$ is made
resonant [see Fig.~\ref{fig:triplet_gen_diagrams}, panels (a) and (b)]. Hence after half a Rabi period $T_1\approx\frac{\pi}{2}\left( \Gamma_P+\frac{4\Gamma_S\Gamma_+}{U}
  \right)^{-1}$
the right QD will be doubly occupied and
the left QD will be empty. Note that $\Gamma_S$ is the CPS rate in a finite Zeeman field. Spurious contributions from
off-resonant terms will limit the maximal achievable double occupancy
to roughly
\begin{align}\label{eq:7}
\mathcal{F}_D\approx 1-8\left( \frac{\Gamma_S}{U}\right)^{2}.
\end{align}
The derivation of Eq.~(\ref{eq:7}) and of the approximate expression for
$T_1$ are given in Appendix~\ref{app:Triplet}.
\begin{figure}[t]
\includegraphics[width=\columnwidth]{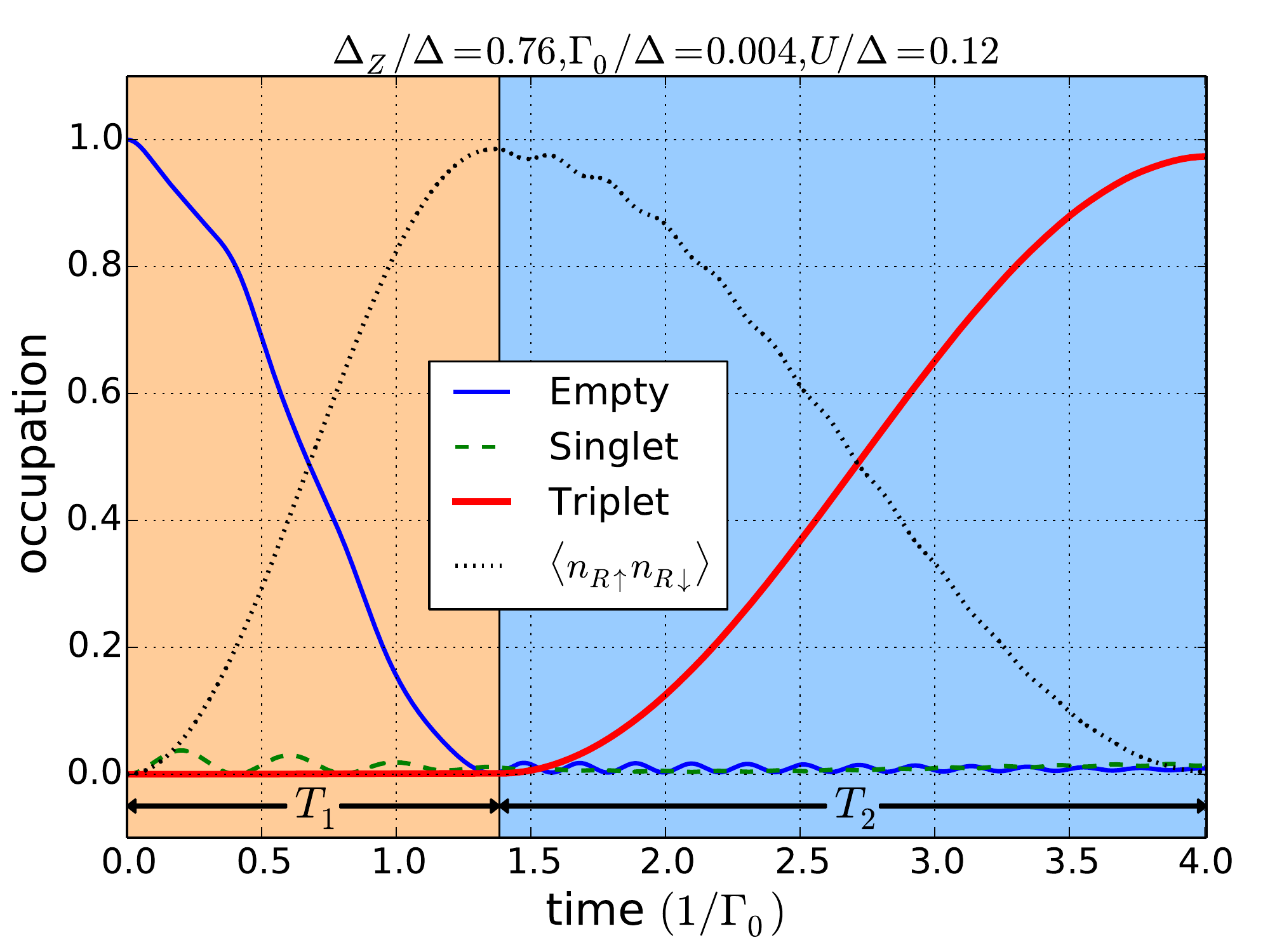}\caption{Dynamics of
  the triplet state generation for $U/\Delta=0.12$,
  $\Delta_Z/\Delta=0.76$ and $\Gamma_0/\Delta=0.004$. The two shaded areas of the graph
  correspond to the two stages of the protocol described in the text
  and are separated by the
  switching of the potential of the right QD from $-U/2$ to $-U$,
  (see also Fig.~\ref{fig:triplet_gen_diagrams}). In the first stage,
  population is transferred from the vacuum (solid (blue) line) to the doubly occupied
  state of the right QD (dotted (black) line). In the second stage, population is transferred
  from the doubly occupied state to the non local triplet (solid thick (red)
  line). A maximal triplet fidelity of $97\%$ is reached at time $T_1+T_2$. Note also
  the presence of a small oscillatory population of the non local singlet state (dashed
  (green) line).
  \label{fig:triplet_dyn}}
\end{figure}
\begin{figure}[ht]
\includegraphics[width=\columnwidth]{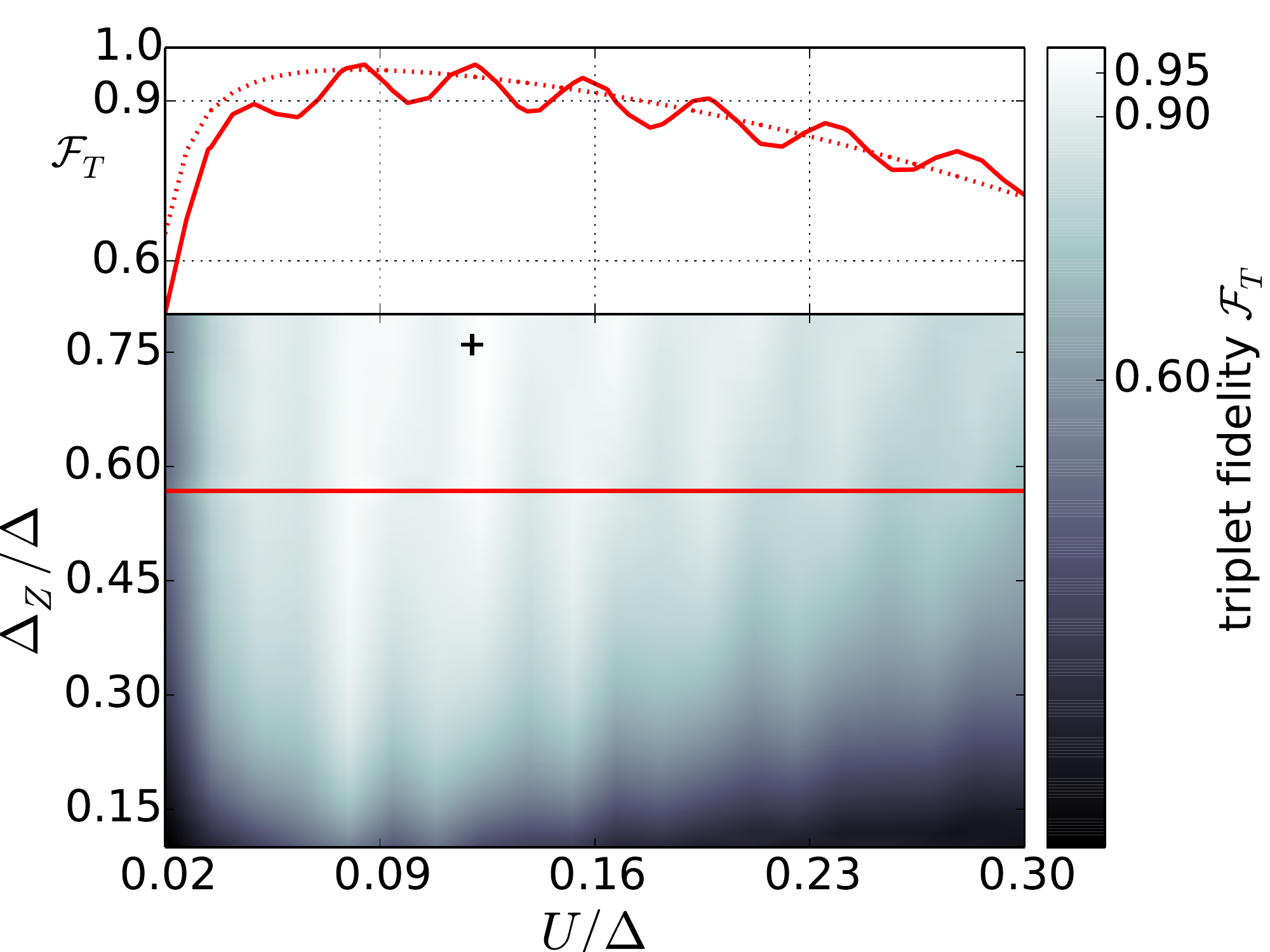}\caption{The contour
  plot shows the fidelity of
  the triplet state generation as a function of on-site interaction
  strength $U$ and Zeeman field $\Delta_Z$ for $\Gamma_0/\Delta=0.004$. The (black) cross indicates the
parameters used in the plot of Fig.~\ref{fig:triplet_dyn}. The solid
(red) line in the upper panel shows the
  fidelity for a fixed value of $\Delta_Z=0.56$ indicated by the solid (red) line in the contour plot. In the regime where
  $\Gamma_0\ll U\ll\Delta_Z\ll\Delta$, the triplet fidelity is well
  approximated by the analytic expression $\mathcal{F}_T\approx
  1-(U/\Delta_Z)^2-8(\Gamma_0/U)^2$ (dotted (red) line). Weak oscillations of the fidelity
as a function of $U$ are clearly visible.\label{fig:contour_plot}}
\end{figure}
At time $T_1$, the right QD is then quickly detuned
further to $\mu_R=-U$. This detunes the pair
tunneling term off resonance and hence
suppress it while making the inter-dot term resonant [see
Fig.~\ref{fig:triplet_gen_diagrams}, panels (c) and (d)]. After another half
Rabi period of $T_2\approx(\pi/2)\left[ 2\left(\Gamma_-^2+\Gamma_+^2\right)\right]^{-1/2}$, a state with a
large non local triplet population is generated on the two QDs. An estimate for the achievable triplet fidelity
in the regime $\Gamma_0\ll U\ll\Delta_Z\ll\Delta$
is given by
\begin{align}
  \mathcal{F}_T\approx \mathcal{F}_D\mathcal{F}_T^{\rm ideal}\approx 1-\left( \frac{U}{\Delta_Z} \right)^2-8\left( \frac{\Gamma_0}{U}\right)^2 .
\end{align}
The expression on the right hand side has a simple interpretation: The term
$8(\Gamma_0/U)^2$ describes the loss of fidelity due to the
competition between resonant pair tunneling and off resonant Cooper
pair splitting in the first stage of the scheme. The term
$(U/\Delta_Z)^2$ describes the loss of fidelity during the second
stage of the scheme due to the residual spin-symmetric
term in $H_D$ that favors singlet pairing and competes with the
spin anti-symmetric term that favors triplet pairing. A more general
analytic expression for the triplet fidelity, which relaxes some of the strong inequality
constraints above, is provided in Appendix~\ref{app:Triplet}.

To confirm the above picture we have
numerically solved the Schr\"odinger equation with the full
Hamiltonian~(\ref{Low_Energy_Hamiltonian}). The results are illustrated in Fig.~\ref{fig:triplet_dyn}, where it is shown that a
triplet state with $97\%$ fidelity can be obtained for parameters
satisfying $\Gamma_0\ll U\ll\Delta_Z\ll\Delta$.

In the simulation, we assume a chemical potential switching
time fast compared to $T_1$ and $T_2$.

Figure~\ref{fig:contour_plot} shows how the maximal triplet fidelity depends on the on-site
interaction strength and Zeeman field. The general trend is well
captured by the analytic approximation~(\ref{eq:1}) (See upper panel
of Fig.~\ref{fig:contour_plot} for a direct comparison). It is
noteworthy that for small values of $U$, the fidelity suppression is
somewhat stronger than predicted by Eq.~(\ref{eq:1}). This together
with the weak oscillations of the fidelity as a
function of $U$ can be attributed to higher order
terms, neglected in the analytic approximation.

In order to experimentally verify the successful generation of a non local triplet and distinguishing it from a non local singlet state, we propose two different schemes. Firstly, the QDs could be attached to
mesoscopic wires forming the inputs of an electronic beam-splitter.
Depending on whether the two interfering electrons form a singlet or a triplet, the sign of the two particle interference term will differ~\cite{Burkard, Samuelsson}.
Secondly, a gate tunable direct inter-dot tunneling term makes it possible to employ the spin-blockade technique
pioneered in~\cite{Ono-2002, Petta-2005}. This enables a mapping from the two spin states onto two distinct
charge states of one of the two QDs. The charge states can then be distinguished via a capacitively coupled rf-single electron transistor device~\cite{Schoelkopf-1998, Gordon}.


\section{Conclusions}
In conclusion, our work represents a first step in
the investigation of the CPS beyond the infinite $U$ limit. We analyze effects of electron-electron
interaction in the CPS in the presence or absence of a magnetic Zeeman
field. We derive an analytic low energy effective Hamiltonian for this
system and identify a novel term that describes an inter-QD interaction
mediated by the superconductor. We make use of this interaction and of
the electrostatic tunability of the QD
levels, to propose a scheme to generate a non local triplet state on
the two QDs. Thereby we extend the capabilities of the
CPS to generate two of the four maximally entangled Bell states with
high fidelities. Experimental investigations testing
the validity of the presented effective low energy model in this novel parameter
regime seem feasible with current technologies and would be a very
useful step towards quantum state engineering with the CPS.


\section*{Acknowledgments}
We would like to thank Christoph Bruder for stimulating discussions.
E.~A. and R.~P.~T. acknowledge financial support from the Swiss National Science Foundation and NCCR Quantum Science and Technology. S.~E.~N. is supported by an Ambizione fellowship from the SNF.
S.~W. acknowledges financial support by the Marie Curie ITN cQOM. T.~L.~S. is supported by the National Research Fund Luxembourg (ATTRACT 7556175).

\appendix

\section{Analytic model of triplet state generation} \label{app:Triplet}
In this appendix, we introduce and analytically solve a simplified
model for the dynamics of the triplet state generation presented in
Section~\ref{Triplet_generation}. We motivate this model using Fig.~\ref{fig:triplet_dyn}, which
shows the population dynamics in the parameter regime suitable for
triplet generation. The crucial observation is that in each of the two stages of
the drive scheme, only three states are significantly populated. More
specifically, in stage I, these states are (i) the vacuum $\ket{V}=\ket{0}_L\ket{0}_R$,
(ii) the doubly occupied state of the right QD $\ket{D}=c_{R+}^{\dagger}c_{R-}^{\dagger}\ket{V}$ and (iii)
the non local singlet state
$\ket{S}=(c_{L+}^{\dagger}c_{R-}^{\dagger}-c_{L-}^{\dagger}c_{R+}^{\dagger})/\sqrt{2}\ket{V}$. In stage
II the three states are (i) the doubly occupied state $\ket{D}$, (ii)
the non local triplet state $\ket{T}=(c_{L+}^{\dagger}c_{R-}^{\dagger}+c_{L-}^{\dagger}c_{R+}^{\dagger})/\sqrt{2}\ket{V}$ and (iii) the non local singlet
$\ket{S}$. This fact suggests that we can approximately neglect the
occupation of all other states and project the system onto three
dimensional subspaces in both stages I and II and match the solutions
at the interface (i.e. at time $T_1$). We proceed by treating the two
stages separately.

\textbf{Stage I:} In the subspace$\{\ket{V},\ket{D},\ket{S}\}$ the Hamiltonian is given by
\begin{align}
H_{\rm I}=
	\begin{pmatrix}
		0 & \Gamma_\text{P} & \sqrt{2}\Gamma_\text{S} \\
		\Gamma_\text{P} & 0 & \sqrt{2}\Gamma_+ \\
		\sqrt{2}\Gamma_\text{S} & \sqrt{2}\Gamma_+ & -U/2 \\
	\end{pmatrix},
\end{align}
where the matrix elements are defined in terms of the rates given in
the main text (See Eqs.~(\ref{eq:5}),~(\ref{gamma_+}) and~(\ref{gamma_S})) and the factors of $\sqrt{2}$ appear because of the
normalization of the singlet state.
We now make use of the fact that in stage I, the QDs are tuned such
that the vacuum state and the
doubly occupied state are resonant with each other while the singlet
is off resonance by $U/2\gg \Gamma_S,\Gamma_+,\Gamma_P$. To carry out
the degenerate perturbation theory, we
switch to a new basis given by the states
\begin{align}
\ket{0}&\equiv\frac{1}{\sqrt{2}}\left( \ket{V}+\ket{D} \right)\\
\ket{1}&\equiv\frac{1}{\sqrt{2}}\left( \ket{V}-\ket{D} \right)\\
\ket{2}&\equiv\ket{S}.
\end{align}
In this new basis, the Hamiltonian takes the form
\begin{align}
H
=
	\begin{pmatrix}
		\Gamma_\text{P} & 0 & \Gamma_\text{S}+\Gamma_\text{+} \\
		0 & -\Gamma_\text{P} &  \Gamma_\text{S}-\Gamma_\text{+}  \\
		 \Gamma_\text{S}+\Gamma_\text{+}  & \Gamma_\text{S}-\Gamma_\text{+}  & -U/2 \\
	\end{pmatrix}.
\end{align}
Treating the off-diagonal terms as perturbation, we find the corrections
to the states $\ket{0},  \ket{1}$ and $\ket{2}$ and the corresponding eigenenergies up to second order:
\begin{align}
	\ket{\tilde 0} &= \frac{\ket{V}+\ket{D}}{\sqrt{2}} + \frac{\Gamma_\text{S}+\Gamma_\text{+}}{U/2+\Gamma_\text{P}}\ket{S},  \\
	\ket{\tilde 1} &= \frac{\ket{V}-\ket{D}}{\sqrt{2}}+\frac{\Gamma_\text{S}-\Gamma_\text{+}}{U/2-\Gamma_\text{P}}\ket{S}, \\
	\ket{\tilde 2} &=  \ket{S}\\
&+\sqrt{2}\left( \frac{\Gamma_\text{S}
                         U/2-\Gamma_\text{+}\Gamma_\text{P}}{\Gamma_\text{P}^2-(U/2)^2}\ket{V}+\frac{\Gamma_\text{+}
                         U/2-\Gamma_\text{S}\Gamma_\text{P}}{\Gamma_\text{P}^2-(U/2)^2}\ket{D}
                         \right)\nonumber
\\
	\tilde E_0 &= \Gamma_\text{P}+\frac{(\Gamma_\text{S}+\Gamma_\text{+})^2}{U/2+\Gamma_\text{P}}, \\
	\tilde E_1 &= -\Gamma_\text{P}+\frac{(\Gamma_\text{S}-\Gamma_\text{+})^2}{U/2-\Gamma_\text{P}}, \\
	\tilde E_2 &= -\frac{U}{2}-\frac{(\Gamma_\text{S}+\Gamma_\text{+})^2}{U/2+\Gamma_\text{P}}-\frac{(\Gamma_\text{S}-\Gamma_\text{+})^2}{U/2-\Gamma_\text{P}}.
\end{align}
Assuming that at time $t=0$ the system is in the vacuum state
$\ket{V}$, we can approximate the state at time $t$ as
\begin{equation}
	\ket{\psi(t)}
=
	\sum_{n=0}^2
	e^{-i\tilde E_nt/\hbar}\braket{\tilde n|V}
	\ket{\tilde n}.
\end{equation}
Hence the probability that the right QD is doubly occupied at time $t$
(equivalent to the fidelity of the doubly occupied state)
is found to be
\begin{multline}\label{D}
	\mathcal{F}^{({\rm I})}_{D}(t)=|\braket{D|\psi(t)}|^2
=
	\frac{1}{1+\frac{8\Gamma_\text{S}^2}{U^2}}
	\left\{
		\sin^2\left[t\left(\Gamma_\text{P}+\frac{4\Gamma_\text{S}\Gamma_\text{+}}{U}\right)\right]
	\right.
\\
	\left.
	-
		\frac{16\Gamma_\text{S}\Gamma_\text{+}}{U^2}
		\sin\left(\Gamma_\text{P}t\right)
		\sin\left[t\left(\frac{U}{2}+6\frac{\Gamma_\text{S}^2+\Gamma_\text{+}^2}{U}\right)\right]
	\right\}.
\end{multline}
The term on the first line of this equation described the leading
order suppression of the fidelity due to the off-resonant transitions
between the vacuum and the singlet while the second line describes
higher order corrections ($\sim \mathcal{O}\left( \Gamma_S\Gamma_+/U^2 \right)$). Physically the latter describe the
second order process where a non local singlet is first created out of
the vacuum and then transferred to a doubly occupied state by the action of the spin-symmetric part of the
inter-dot tunneling Hamiltonian. Because the amplitude of this process
adds coherently, it leads to small amplitude oscillations of the
fidelity at a frequency of the order of charging energy
$U$. Neglecting these small oscillations and expanding to leading
order, we obtain the
estimates for the optimal double occupancy time $T_1\approx
\frac{\pi}{2} \left(\Gamma_P+\frac{4\Gamma_S \Gamma_+}{U}\right)^{-1}$
as well as the maximal fidelity of double occupancy
$\mathcal{F}_D\approx 1-8(\Gamma_S/U)^2$ given in Eq.~(\ref{eq:7}).
Using the perturbative approach, we note also that the probability for the system to be found in the vacuum state $\ket{V}$ at time $t$ is 
\begin{multline}\label{2}
	\mathcal{F}^{({\rm I})}_{V}(t)
=
	\frac{1}{1+\frac{8\Gamma_\text{S}^2}{U^2}}
	\left\{
		\cos^2\left[t\left(\Gamma_\text{P}+\frac{4\Gamma_\text{S}\Gamma_+}{U}\right)\right]
	\right.
\\
	\left.
	+
		\frac{16\Gamma_\text{S}^2}{U^2}
		\cos\left(\Gamma_\text{P}t\right)
		\cos\left[t\left(\frac{U}{2}+6\frac{\Gamma_\text{S}^2+\Gamma_+^2}{U}\right)\right]
	\right\},
\end{multline}
and the the probability for the system to be found in the singlet state $\ket{S}$ at time $t$ is given by
\begin{align} \label{3}
&	\mathcal{F}^{({\rm I})}_{S}(t)
=
	\frac{2}{U^2+8\Gamma_\text{S}^2}
	\left\{4\Gamma_S^2+
		\Gamma_\text{S}^2\sin^2\left[t\left(\Gamma_\text{P}+\frac{4\Gamma_\text{S}\Gamma_+}{U}\right)\right]
	\right.
\\
&	\left.\qquad\qquad\qquad
	+
		4\Gamma_\text{S}^2
		\left( 
			\cos^2\left[t\left(\Gamma_\text{P}+\frac{4\Gamma_\text{S}\Gamma_+}{U}\right)\right]\nonumber
\right.\right.
\\
&\left.\left. \qquad\qquad\qquad\qquad
		-
		2\cos\left(\Gamma_\text{P}t\right)
		\cos\left[t\left(\frac{U}{2}+6\frac{\Gamma_\text{S}^2+\Gamma_+^2}{U}\right)\right]
		\right)
	\right\}\nonumber
\end{align}

\textbf{Stage II:} The analysis of stage II is very similar. In this
stage the relevant states form the subspace $\{\ket{D}, \ket{T},
\ket{S}\}$. The resulting three level Hamiltonian is given by
\begin{equation}
H=
	\begin{pmatrix}
		-U & \sqrt{2}\Gamma_- & \sqrt{2}\Gamma_+ \\
		\sqrt{2}\Gamma_- & -U & 0 \\
		\sqrt{2}\Gamma_+ & 0 & -U \\
	\end{pmatrix},
\end{equation}
where $\Gamma_-$ was defined in Eq.~(\ref{gamma_-}). Owing to the
threefold degeneracy of the bare states, this Hamiltonian
can easily be diagonalized. The eigenstates are
given by
\begin{align}
	\ket{\psi_0}
=
	\frac{1}{\sqrt{\Gamma_\text{+}^2+\Gamma_-^2} }
	\left(
		-\Gamma_+\ket{T}+\Gamma_-\ket{S}
	\right),
\end{align}
\begin{align}
	\ket{\psi_1}
=
	\frac{1}{\sqrt{2\left(\Gamma_\text{+}^2+\Gamma_-^2 \right)}}
	\left(
		- \sqrt{\Gamma_+^2+\Gamma_-^2}\ket{D}+\Gamma_-\ket{T}+\Gamma_\text{+}\ket{S}
	\right),
\end{align}
\begin{align}
	\ket{\psi_2}
=
	\frac{1}{\sqrt{2\left(\Gamma_\text{+}^2+\Gamma_-^2 \right)}}
	\left(
		\sqrt{\Gamma_+^2+\Gamma_-^2}\ket{D}+\Gamma_-\ket{T}+\Gamma_+\ket{S}
	\right).
\end{align}
The corresponding eigenenergies are
\begin{align}
E_0&=-U,\\
E_1&=-U-\sqrt{2\left(\Gamma_\text{+}^2+\Gamma_-^2\right)},\\
E_2&=-U+\sqrt{2\left(\Gamma_\text{+}^2+\Gamma_-^2\right)}.
\end{align}

To account for an imperfect state preparation after stage I,
we consider and initial state for stage II of the form
$a\ket{D}+b\ket{S}$ with $|a|^2+|b|^2=1$. With this, the
probability of finding the system in the triplet state $\ket{T}$
at time $t$ is given by
\begin{multline} \label{eq:8}
	\mathcal{F}^{({\rm II})}_{T}(t)
=
\frac{\Gamma_-^2}{\left(\Gamma_+^2+\Gamma_-^2\right)^2}\times
	\left\{	|a|^2\left(\Gamma_+^2+\Gamma_-^2\right)\sin^2\left(t\sqrt{2\left(\Gamma_+^2+\Gamma_-^2\right)}\right)
	\right.
\\
\left.	+|b|^2\Gamma_+^2\left[1-2\cos\left(t\sqrt{2\left(\Gamma_+^2+\Gamma_-^2\right)}\right)+\cos^2\left(t\sqrt{2\left(\Gamma_+^2+\Gamma_-^2\right)}\right)\right]
\right\}.
\end{multline}
Note that because only
the initial state probabilities $|a|^2$ and $|b|^2$ appear in
Eq.~(\ref{eq:8}), there are no interference terms between stage I and
II in the present approach. Thus we can simply obtain the triplet
fidelity by multiplying the fidelities of the double
occupancy~(\ref{D}) and the {\em ideal} triplet fidelity obtained
from~(\ref{eq:8}) by setting $a=1$ and $b=0$. The latter is given by
\begin{align}
\mathcal{F}_{T}^{\rm ideal}(t)=\frac{1}{1+\left( \frac{\Gamma_+}{\Gamma_-} \right)^2}\sin^2\left( t\sqrt{2\left( \Gamma_+^2+\Gamma_-^2 \right)} \right),
\end{align}
from which we immediately obtain an estimate for the ideal duration of
stage II: $T_2\approx(\pi/2)\left[
  2\left(\Gamma_-^2+\Gamma_+^2\right)\right]^{-1/2}$. Combining these
expressions and expanding to leading order in the regime where
$\Gamma_0\ll U\ll\Delta_Z\ll\Delta$, we obtain the expression for the triplet
fidelity of Eq.~(\ref{eq:1}).

\begin{figure}[t]
\includegraphics[width=\columnwidth]{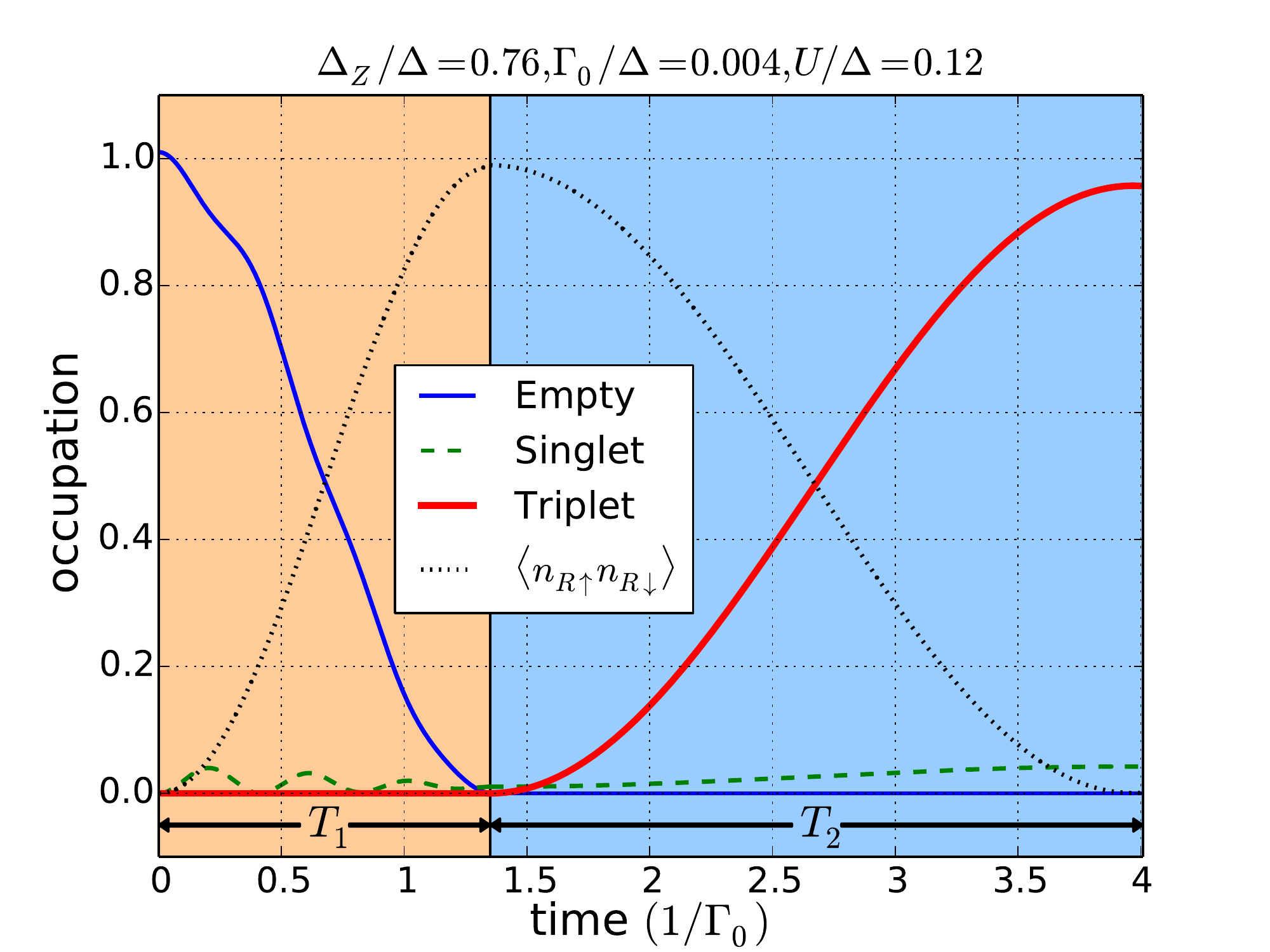}\caption{Analytical prediction for the dynamics of
  the triplet state generation for $U/\Delta=0.12$,
  $\Delta_Z/\Delta=0.76$ and $\Gamma_0/\Delta=0.004$. This result
  agrees qualitatively with the full numerics shown in Fig.~\ref{fig:triplet_dyn}.
  \label{fig:analytical}}
\end{figure}

Using this analytic model, we also note that the probability for the system to be found in the doubly occupied state $\ket{D}$ at time $t$ is given by
\begin{multline}\label{5}
	\mathcal{F}^{({\rm II})}_{D}(t)
=
	|a|^2
	\cos^2\left(t\sqrt{2\left(\Gamma_+^2+\Gamma_-^2\right)}\right)
\\
+
	\frac{|b|^2}{1+\left(\frac{\Gamma_-}{\Gamma_+}\right)^2}\sin^2\left(t\sqrt{2\left(\Gamma_+^2+\Gamma_-^2\right)}\right),
\end{multline}
while the probability for the system to be found in the singlet state $\ket{S}$ at time $t$ is given by
\begin{multline}\label{6}
	\mathcal{F}^{({\rm II})}_{S}(t)
=
	\frac{\Gamma_+^2}{\left(\Gamma_+^2+\Gamma_-^2\right)^2}\times
	\left\{	|a|^2\left(\Gamma_+^2+\Gamma_-^2\right)\sin^2\left(t\sqrt{2\left(\Gamma_+^2+\Gamma_-^2\right)}\right)
	\right.
\\
\left.	+|b|^2\Gamma_+^2\left[\frac{\Gamma_-^4}{\Gamma_+^4}+\frac{2\Gamma_-^2}{\Gamma_+^2}\cos\left(t\sqrt{2\left(\Gamma_+^2+\Gamma_-^2\right)}\right)+
\right.
\right.
\\
\left.
\left. \cos^2\left(t\sqrt{2\left(\Gamma_+^2+\Gamma_-^2\right)}\right)\right]
\right\}.
\end{multline}
We have plotted in Fig.~\ref{fig:analytical} the occupation of the different states in stage I and in stage II using the analytical results obtained, Eq.~(\ref{D}-\ref{3}), Eq.~(\ref{eq:8}) and Eq.~(\ref{5}-\ref{6}). This can be compared with Fig.~\ref{fig:triplet_dyn}.


\section{The Schrieffer-Wolff transformation with finite on-site
  repulsion and Zeeman field} \label{App:SW}
In this Appendix, we present the SW transformation we have used in order to eliminate the first order tunneling term appearing in the original Hamiltonian $H$, Eq.~(\ref{Original_Hamiltonian}), thus leading (after neglecting higher than second order terms in the tunneling amplitude) to the effective low energy Hamiltonian $H_\text{LE}$, Eq.~(\ref{Low_Energy_Hamiltonian}). 

The SW transformation~\cite{SchriefferWolff} is a unitary transformation, $U\equiv e^S$. After transforming $H$ we obtain
\begin{equation}
	\tilde{H}
=
	H
+
	[S,H]
+
	\frac{1}{2!}
	\left[
		S,
		[S,H]
	\right]
+
	\ldots
+
	\frac{1}{n!}
	\left[
		S
		\left[
		S,
			\left[
				\ldots
				[S,H]
				\ldots
			\right]
		\right]
	\right]
+
	\ldots
			\label{trans_Hamil_U}
\end{equation}
We choose the generator $S$ of the canonical transformation such that it eliminates the perturbation to first order in $w_\lambda$, i.e. 
\begin{equation}
	[S,H_\text{BCS}+H_\text{QDs}]=-K. \label{condition1_U}
\end{equation}
Then, the transformed Hamiltonian becomes, 
\begin{equation}
	\tilde{H}
=
	H_\text{BCS}+H_\text{QDs}
+
	\frac{1}{2}
	[S,K]
+
	\mathcal{O}(w_\lambda^3).
\end{equation}
Keeping terms up to second order in $w_\lambda/\Delta$, we find our low energy Hamiltonian,
\begin{equation}
	H_\text{LE}
=
	H_\text{BCS}+H_\text{QDs}
+
	H_\text{SW},
\end{equation}
where we have defined $H_\text{SW} \equiv [S,K]/2$.
The solution of Eq.~(\ref{condition1_U}) is given by~\cite{Salomaa}
\begin{equation} \label{Our_S}
\begin{split}
	S
=
	&\sum_\lambda \sum_k \sum_\sigma
	w_{\lambda k}
	\left[
		\left(
			\frac{1-n_{\lambda \bar{\sigma}}}{E_k-\epsilon_{\lambda,\sigma}}
		+
			\frac{n_{\lambda \bar{\sigma}}}{E_k-\epsilon_{\lambda,\sigma}-U_\lambda}
		\right)
		u_k \alpha_{k\sigma}^\dagger c_{\lambda \sigma}
	\right.
\\&
	\left.
		+
		\sigma
		\left(
			\frac{1-n_{\lambda \sigma}}{E_k+\epsilon_{\lambda,\bar{\sigma}}}
		+
			\frac{n_{\lambda \sigma}}{E_k+\epsilon_{\lambda,\bar{\sigma}}+U_\lambda}
		\right)
		v_k \alpha_{k \sigma} c_{\lambda \bar{\sigma}}
	\right]
-
	{\rm h.c.},
\end{split}
\end{equation}
as can be easily verified by substitution. Using this generator, one can calculate $H_\text{SW}$. This can be done quite generally, for different values of $\Delta, \Delta_Z, U_\lambda$ and $K_BT$ (as long as one keeps in mind that keeping the leading order term in the SW transformation is justified for $\Delta \gg K_BT$). However, as we are interested in the regime where the superconducting gap $\Delta$ is much larger than the thermal energy, we furthermore assume that the superconductor is at zero temperature (i.e. we eliminate the quasi particle $\alpha_{k\sigma}$ degrees of freedom by taking the expectation value of $H_\text{SW}$ in a state with no quasi-particles). After integrating out the $k$ dependence using the assumption $\Delta>U_\lambda+\Delta_Z/2$, one obtains an effective low energy Hamiltonian,

\begin{equation} \label{H_full}
	H_\text{LE}
=
	H_\text{QDs}
+
	H_S
+
	H_P
+
	H_D,
\end{equation}
where $H_\text{QDs}$ is the QDs Hamiltonian appearing in Eq.~(\ref{eq:3}), and the other terms are given by
\begin{widetext}
\begin{align}
	H_S
&=
	\frac{\pi \rho_0 w_{L}w_{R}}{2}
\left\{
\left(
			\frac{1}
			{\sqrt{1-\left(\frac{\Delta_Z}{2\Delta}\right)^2}}
			(2-n_{L -}-n_{R+})
	+
			\frac{1}
			{\sqrt{1-\left(\frac{\Delta_Z/2+U_L}{\Delta}\right)^2}}
			n_{L -}
	+
			\frac{1}
			{\sqrt{1-\left(\frac{\Delta_Z/2-U_R}{\Delta}\right)^2}}
			n_{R+}
	\right)
	c_{L+}^\dagger c_{R-}^\dagger 
\right.
\nonumber\\&\left.
	-
	\left(
			\frac{1}
			{\sqrt{1-\left(\frac{\Delta_Z}{2\Delta}\right)^2}}
			(2-n_{R-}-n_{L+})
	+
			\frac{1}
			{\sqrt{1-\left(\frac{\Delta_Z/2-U_L}{\Delta}\right)^2}}
			n_{L+}
	+
			\frac{1}
			{\sqrt{1-\left(\frac{\Delta_Z/2+U_R}{\Delta}\right)^2}}
			n_{R-}
	\right)
	c_{L-}^\dagger c_{R+}^\dagger  
\right\} +{\rm h.c.},
\\
	H_P
&=
	\frac{\pi \rho_0 }{2}
\left\{
		\left(
			\frac{1}
			{\sqrt{1-\left(\frac{\Delta_Z}{2\Delta}\right)^2}}
		+
			\frac{1/2+\frac{1}{\pi}\tan^{-1}\left(\frac{\frac{\Delta_Z/2+ U_R}{\Delta}}{\sqrt{1-\left(\frac{\Delta_Z/2+ U_R}{\Delta}\right)^2}}\right)}
			{\sqrt{1-\left(\frac{\Delta_Z/2+U_R}{\Delta}\right)^2}}
		+
			\frac{1/2-\frac{1}{\pi}\tan^{-1}\left(\frac{\frac{\Delta_Z/2- U_R}{\Delta}}{\sqrt{1-\left(\frac{\Delta_Z/2- U_R}{\Delta}\right)^2}}\right)}
			{\sqrt{1-\left(\frac{\Delta_Z/2-U_R}{\Delta}\right)^2}}
		\right)
		w_{R }^2  c_{R+}^\dagger c_{R -}^\dagger
\right.
\nonumber\\&\left.
		+
		\left(
			\frac{1}
			{\sqrt{1-\left(\frac{\Delta_Z}{2\Delta}\right)^2}}
		+
			\frac{1/2+\frac{1}{\pi}\tan^{-1}\left(\frac{\frac{\Delta_Z/2+U_L}{\Delta}}{\sqrt{1-\left(\frac{\Delta_Z/2+U_L}{\Delta}\right)^2}}\right)}
			{\sqrt{1-\left(\frac{\Delta_Z/2+U_L}{\Delta}\right)^2}}
		+
			\frac{1/2-\frac{1}{\pi}\tan^{-1}\left(\frac{\frac{\Delta_Z/2-U_L}{\Delta}}{\sqrt{1-\left(\frac{\Delta_Z/2-U_L}{\Delta}\right)^2}}\right)}
			{\sqrt{1-\left(\frac{\Delta_Z/2-U_L}{\Delta}\right)^2}}
		\right)
		w_{L}^2 c_{L+}^\dagger c_{L-}^\dagger 
\right\} +{\rm h.c.},
\\
	H_D
&=
	\frac{\pi \rho_0 w_L w_R }{2}
\left\{
	\left(
	\frac{1}
	{\sqrt{\left(\frac{2\Delta}{\Delta_Z}\right)^2-1}}
	(2-n_{L-}-n_{R-})
	+
	\frac{1}
	{\sqrt{\left(\frac{\Delta}{\Delta_Z/2+U_L}\right)^2 - 1}}
	n_{L-}
	+
	\frac{1}
	{\sqrt{\left(\frac{\Delta}{\Delta_Z/2+U_R}\right)^2 - 1}}
	n_{R-}
	\right)
	c_{L+} c^\dagger_{R+}
\right.
\nonumber\\&
		\left.
	-
	\left(
	\frac{1}
	{\sqrt{\left(\frac{2\Delta}{\Delta_Z}\right)^2-1}}
	(2-n_{L+}-n_{R+})
	+
	\frac{{\rm sgn}\left(\Delta_Z/2-U_R\right)}
	{\sqrt{\left(\frac{\Delta}{\Delta_Z/2-U_R}\right)^2 - 1}}
	n_{R+}
	+
	\frac{{\rm sgn}\left(\Delta_Z/2-U_L\right)}
	{\sqrt{\left(\frac{\Delta}{\Delta_Z/2-U_L}\right)^2 - 1}}
	n_{L+}
	\right)
	c_{L-} c_{R-}^\dagger 
	\right\}
+{\rm h.c.},
\end{align}
\end{widetext}
where $H_S$ contains the standard Cooper pair splitting process which,
in contrast to the Cooper pair splitting described using the Coulomb
blockade approximation, depends now on the occupation of the QDs via
the number operators. $H_p$ describes the tunneling of a Cooper pair
onto a single QD, and $H_D$ describes and effective inter-dot
tunneling processes which is mediated via the superconductor. The
processes described in $H_P$ and in $H_D$ are obviously not accounted
for in the standard Coulomb blockade approximation. Assuming
$w_L=w_R=w$ and $U_L=U_R=U$, one can obtain the reduced form  of the
low energy Hamiltonian which is given in the main text,
Eq.~(\ref{Low_Energy_Hamiltonian}). In the zero field limit these
expressions further simplify and are provided here for
completeness. They read:
\begin{align}
&H_S^{\Delta_Z\rightarrow 0} =\frac{\Gamma_0}{2}\left\{ \left( 2-n_{L-}^{}-n_{R+}^{}
  +\frac{n_{L-}+n_{R+}}{\sqrt{1-\left(\frac{U}{\Delta}\right)^2}}
  \right)c_{L+}^{\dagger}c_{R-}^{\dagger}\right.\nonumber\\
&-\left.\left( 2-n_{L+}^{}-n_{R-}^{}
  +\frac{n_{L+}+n_{R-}}{\sqrt{1-\left(\frac{U}{\Delta}\right)^2}}
  \right)c_{L-}^{\dagger}c_{R+}^{\dagger}\right\}+{\rm h.c.}
\end{align}

\begin{align}
H_P^{\Delta_Z\rightarrow 0}&=\frac{\Gamma_0}{2}\left( 1+\frac{1+\frac{2}{\pi}\tan^{-1}\left(
     \frac{1}{\sqrt{\left( \frac{\Delta}{U} \right)^2-1}}
     \right)}{\sqrt{1-\left( \frac{U}{\Delta} \right)^2}}
     \right)\sum_{\alpha=L,R}c_{\alpha+}^{\dagger}c_{\alpha-}^{\dagger}
    +{\rm h.c.}
\end{align}
and
\begin{align}
H_D^{\Delta_Z\rightarrow 0}&=\frac{\Gamma_0}{2}\frac{1}{\sqrt{\left(
                             \frac{\Delta}{U} \right)^2-1}}\sum_{\sigma=\pm}
                             \left( n_{L\sigma}+n_{R\sigma}
                             \right)c_{L\bar\sigma}^{}c_{R\bar\sigma}^{\dagger}+{\rm
                             h.c.}
\end{align}

Further insight into the effective Hamiltonian in Eq.~(\ref{H_full}) can be gained by examining the limiting case of an infinite superconducting gap. In that limit, important for transport processes involving Andreev reflection, effective Hamiltonians for proximised QDs have already been introduced in the literature~\cite{Eldridge, Sothmann, Droste}. Taking this limit in Eq.~(\ref{H_full}), one obtains 
\begin{multline}
	H_\text{LE}^{\Delta \rightarrow \infty} = H_\text{QDs}
        +\pi\rho_0\omega_L\omega_R \left(c_{L+}^\dagger
          c_{R-}^\dagger-c_{L-}^\dagger c_{R+}^\dagger + {\rm h.c.}\right)
\\
	+\pi\rho_0\left(\omega_R^2 c_{R+}^\dagger c_{R-}^\dagger +
          \omega_L^2 c_{L+}^\dagger c_{L-}^\dagger+{\rm h.c.}\right).
\end{multline}
This result agrees with~\cite{Sothmann} for example.

\bibliographystyle{apsrev}
 \bibliography{paper}

\end{document}